\documentclass[aps,prl,twocolumn,
superscriptaddress]{revtex4}
\usepackage{amssymb,amsmath,graphicx,color,bbold}
\begin{document}

\title{Consistent massive graviton on arbitrary backgrounds}

\author{Laura Bernard}
\affiliation{UPMC-CNRS, UMR7095, Institut d'Astrophysique de Paris, GReCO, 98bis boulevard Arago, F-75014 Paris, France}

\author{C\'edric Deffayet}
\affiliation{UPMC-CNRS, UMR7095, Institut d'Astrophysique de Paris, GReCO, 98bis boulevard Arago, F-75014 Paris, France}
\affiliation{IHES, Le Bois-Marie, 35 route de Chartres, F-91440 Bures-sur-Yvette, France}

\author{Mikael von Strauss}
\affiliation{UPMC-CNRS, UMR7095, Institut d'Astrophysique de Paris, GReCO, 98bis boulevard Arago, F-75014 Paris, France}

\begin{abstract}
We obtain the fully covariant linearized field equations for the metric perturbation in the de Rham-Gabadadze-Tolley (dRGT) ghost free massive gravities.  For a subset of these theories, we show that the nondynamical metric
that appears in the dRGT setup can be completely eliminated leading to the theory of a massive graviton moving in a single metric. This has a mass term which contains nontrivial contributions of the space-time curvature. We show further how five covariant constraints can be obtained including one which leads to the tracelessness of the graviton on flat space-time and removes the Boulware-Deser ghost. 
The five constraints are obtained for a background metric which is arbitrary, i.e.~which does not have to obey the background field equations.
\end{abstract}

\maketitle

\newcommand{\ba}{\begin{eqnarray}}
\newcommand{\ea}{\end{eqnarray}}
\newcommand{\be}{\begin{equation}}
\newcommand{\ee}{\end{equation}}
\newcommand{\thetavierbein}{E}
\newcommand{\fvierbein}{L}
\newcommand{\fmoinsunvierbein}{\ell}
\newcommand{\E}{F}
\newcommand{\cc}{c}
\newcommand{\w}{w}
\newcommand{\wb}{\bar{w}}
\newcommand{\W}{W}

\newcommand{\gmn}{g_{\mu\nu}}
\newcommand{\fmn}{f_{\mu\nu}}
\newcommand{\bgmn}{\bar g_{\mu\nu}}
\newcommand{\bfmn}{\bar f_{\mu\nu}}
\newcommand{\pmn}{P_{\mu\nu}}
\newcommand{\smn}{S_{\mu\nu}}
\newcommand{\rmn}{R_{\mu\nu}}
\newcommand{\amn}{A_{\mu\nu}}
\newcommand{\hmn}{h_{\mu\nu}}
\newcommand{\Vmn}{V_{\mu\nu}}
\newcommand{\Gmn}{\mathcal{G}_{\mu\nu}}
\newcommand{\emn}{\eta_{\mu\nu}}
\newcommand{\md}{\mathrm{d}}
\newcommand{\humn}{h^{\mu\nu}}

\newcommand{\beqn}{\begin{eqnarray}}
\newcommand{\eeqn}{\end{eqnarray}}
\newcommand{\td}{\mathrm{d}}
\newcommand{\tD}{\mathcal{D}}
\newcommand{\Lag}{\mathcal{L}}
\newcommand{\GE}{\mathcal{G}}
\newcommand{\p}{\partial}
\newcommand{\dd}{\mathrm{d}}
\newcommand{\nn}{\nonumber}
\newcommand{\Tr}{\mathrm{Tr}}
\newcommand{\A}{\mathbb{A}}
\newcommand{\B}{\mathbb{B}}
\newcommand{\X}{\mathbb{X}}
\newcommand{\Pmn}{P^\mu_{\ph\mu\nu}}
\newcommand{\Smn}{S^\mu_{\ph\mu\nu}}
\def\ph{\phantom}
\newcommand{\Jaa}{\nabla_{\rho}\nabla_{\sigma}\,h^{\rho\sigma}}
\newcommand{\Abb}{[S^3]^{\rho\sigma}\,[S^3]^{\mu\nu}\,\nabla_{\rho}\nabla_{\sigma}\,h_{\mu\nu}}
\newcommand{\ccc}{u}
\newcommand{\ddd}{v}

\newcommand{\gf}{{\mathfrak{F}}}
\newcommand{\scah}{\aleph}

Lately there has been a renewal of interest in massive gravity with interesting applications to cosmology (see e.g.~\cite{reviews} for reviews).
The only consistent linear theory for a massive graviton on flat space-time has been known for a long time since the work of Fierz and Pauli \cite{Fierz:1939ix}. It propagates 5 degrees of freedom of positive energy, those of a transverse, traceless, symmetric, two times covariant tensor $h_{\mu \nu}$. It can easily be extended to an Einstein space-time background keeping the same number of propagating polarizations \cite{Higuchi:1986py,Buchbinder:1999ar}. However, a similar theory for an arbitrary background metric has not been written so far. A starting point to do so is the set of fully non linear theories formulated by de Rham, Gabadadze and Tolley (dRGT in the following) \cite{deRham:2010kj}. Such a theory was shown to contain only 5 dynamical degrees of freedom \cite{deRham:2010kj,Hassan:2011ea} and hence is devoid from a pathology long thought unavoidable: the presence of an extra ghostlike sixth degree of freedom in a generic non linear extension of Fierz-Pauli theory first discussed by Boulware and Deser in the seminal paper \cite{Boulware:1973my}. Hence, {\it a priori} one should be able to extract from the dRGT family a consistent linear theory for a massive graviton moving on a large class of metrics~\footnote{Note that several potential problems of massive gravity have been, and still are, debated in the literature. We refer the reader to Ref.~\cite{reviews} for discussions of these issues, some of which (such as the meaning and consequences of superluminality which was shown to arise for some specific backgrounds) are equally relevant for the linear theory extracted here.}. However, the dRGT theory is formulated using two metrics, a dynamical one, called $\gmn$ in the following, and a non dynamical one, usually taken to be flat, called here $\fmn$. Hence, the expectation is that when one linearizes the dRGT field equations one will obtain a theory for a massive graviton moving in a space-time endowed with two background metrics which has various drawbacks (see e.g.~Ref.~\cite{defjac}).
Moreover, this linearization is not easy, in part because dRGT theories involve a matrix square root $S$ of the tensor $\gf$ both defined by
\ba \label{Sdef} 
S^\mu_{\hphantom{\mu} \sigma}S^\sigma_{\hphantom{\sigma} \nu} = g^{\mu \sigma} f_{\sigma \nu} = \gf^{\mu}_{\hphantom{\mu} \nu}\, .
\ea
Lastly, it may not be easy to show that the obtained theory contains the correct number of degrees of freedom without using the elegant but involved proof obtained for the non linear theories \cite{Hassan:2011ea}. Here, we will overcome these various difficulties and show how to obtain from dRGT models a fully covariant theory for a massive graviton moving in a single, totally arbitrary, metric (hence eliminating the need for the nondynamical metric). We will also show how, for such a theory, one can obtain 5 covariant constraints, including one which leads to the tracelessness of the graviton on flat space-time and removes the Boulware-Deser ghost~\footnote{Note that 
a covariant constraint counting can be obtained in the vierbein and non perturbative formulation \cite{Hinterbichler:2012cn} of dRGT theories \cite{Deffayet:2012nr,Deser:2014hga}. However this formulation is not fully equivalent to the metric formulation \cite{Deffayet:2012nr,Banados:2013fda}.}. This last constraint involves combinations of the curvature of the background metric which become trivial when this metric describes a flat space-time or a more general Einstein space-time. 

Our starting point is the set of massive gravity theories defined by the following action in {\it four} dimensions \cite{deRham:2010kj,Hassan:2011ea}
\be\label{SdRGT} 
S_{g,m} = M_g^2 \int\td^4 x \sqrt{|g|} \left[R(g) -  2m^2 \sum_{n=0}^3\beta_ne_n(S)\right],
\ee
the $\beta_n$ being dimensionless parameters, and $e_n(S)$ the $n$'th order elementary symmetric polynomial of the eigenvalues of its matrix argument $S$. One has in particular $e_0=1$ and $e_1= \Tr[S]$, where here and henceforth $\Tr[X]=X^\rho_{~\rho}$ indicates a matrix trace operation and we do not write out any more the functional dependence of the $e_n$ when they depend only on $S$ (i.e.~it is to be understood that $e_n \equiv e_n(S)$).
The 
$e_n$  can be constructed iteratively (with $e_0=1$) from  the relation
\be\label{edef}
e_n=-\frac{1}{n}\sum_{k=1}^{n}(-1)^k\Tr[S^k]\,e_{n-k}\,,\qquad n\geq1\,,
\ee
where $S^k$ is the k-th power of the tensor $S^\mu_{\hphantom{\mu} \nu}$ (considered as a matrix), and $S^0$ is just the identity.
The field equations deriving from the action (\ref{SdRGT}) for the dynamical metric $\gmn$ are 
\be\label{g_eom1_first}
E_{\mu\nu} \equiv \mathcal{G}_{\mu\nu}+m^2\,V_{\mu\nu}=0\,,
\ee
where ${\cal G}_{\mu \nu}$ is the Einstein tensor built from the metric $\gmn$, and $V_{\mu\nu}$ is given by
\be\label{Vmn1}
V_{\mu\nu}=g_{\mu\rho}\sum_{n=0}^3\sum_{k=0}^n(-1)^{n+k}\beta_n[S^{n-k}]^\rho_{\ph\rho\nu}\,e_k \,.
\ee
The next step in our derivation consists of linearizing these field equations around a background solution for the dynamical metric $\gmn$, calling $h_{\mu \nu}$ the small perturbation of this metric. This may seem at first sight an easy task, however, it is not because it involves in general computing the variation at first order in $h_{\mu \nu}$ of the matrix square root $S$. 
This variation $\delta S$ obeys,  (with obvious notations) as seen from (\ref{Sdef}),
\be \label{SylvdS}
S^\mu_{\hphantom{\mu} \nu} \left(\delta S\right)^\nu_{\hphantom{\nu} \sigma}+ \left(\delta S\right)^\mu_{\hphantom{\mu} \nu}S^\nu_{\hphantom{\nu} \sigma}=\delta \gf^\mu_{\hphantom{\mu} \sigma} \,,
\ee
which is a special kind of Sylvester matrix equation, where the right hand side is easy to get in terms of $h_{\mu \nu}$ from the definition of $\gf$. It is known that this equation has a unique solution for  $\delta S$ if and only if the spectra of $S^\mu_{\hphantom{\mu} \nu}$ and $-S^\mu_{\hphantom{\mu} \nu}$ do no intersect (which is generically the case here). In this case, one can express the solution for $\delta S$ linearly in terms of $\delta \gf$ \cite{Sylvesterpoly}.
Using our own derivation \cite{Bernard:2015mkk} which is more convenient for our purpose here, we obtain 
\begin{widetext}
\be\label{d_Sm}
\begin{aligned}
\dfrac{{\delta S}^{\lambda}_{\hphantom{\lambda}\mu}}{\delta g_{\rho\sigma}}=\quad &
\dfrac{1}{2}\,g^{\nu \lambda} \Bigl[\, e_{4}\,c_{1}\,\Bigl(
\delta_{\nu}^{\rho}\delta^{\sigma}_{\mu}+\delta_{\nu}^{\sigma}\delta_{\mu}^{\rho}-g_{\mu\nu}g^{\rho\sigma}
\Bigr) + e_{4}\,c_{2}\,\Bigl(
S_{\nu}^{\rho}\delta^{\sigma}_{\mu}+S_{\nu}^{\sigma}\delta_{\mu}^{\rho}-S_{\mu\nu}g^{\rho\sigma}
-\gmn S^{\rho\sigma}
\Bigr) - e_{3}\,c_{1}\,\Bigl(
\delta_{\nu}^{\rho}S^{\sigma}_{\mu}+\delta_{\nu}^{\sigma}S_{\mu}^{\rho} \Bigr) \\[1pt]
& +
e_{4}\,c_{3}\,\Bigl[\delta^{\sigma}_{\mu}[S^{2}]_{\nu}^{\rho} 
+\delta_{\mu}^{\rho}[S^{2}]_{\nu}^{\sigma}-g^{\rho\sigma}[S^{2}]_{\mu\nu}+\delta_{\nu}^{\rho}[S^{2}]^{\sigma}_{\mu}+\delta_{\nu}^{\sigma}[S^{2}]_{\mu}^{\rho}-g_{\mu\nu}[S^{2}]^{\rho\sigma}
\Bigr] +\left(e_{2}\,c_{1}-e_{4}\,c_{3}+e_{3}\,c_{2}\right)\,S_{\mu\nu}S^{\rho\sigma} 
\\[1pt]
&
- e_{3}\,c_{2}\,\Bigl(
S_{\nu}^{\rho}S^{\sigma}_{\mu}+S_{\nu}^{\sigma}S_{\mu}^{\rho} \Bigr)
-e_{3}\,c_{3}\,\Bigl(
S^{\sigma}_{\mu}[S^{2}]_{\nu}^{\rho}+S_{\mu}^{\rho}[S^{2}]_{\nu}^{\sigma}+S_{\nu}^{\rho}[S^{2}]^{\sigma}_{\mu}+S_{\nu}^{\sigma}[S^{2}]_{\mu}^{\rho}
\Bigr) \\[1pt]
& +\left(e_{3}\,c_{3}-e_{1}\,c_{1}\right)\,\left(
S^{\rho\sigma}[S^{2}]_{\mu\nu}+S_{\mu\nu}[S^{2}]^{\rho\sigma} \right)
 - \left(c_{1}-e_{2}\,c_{3}\right)\,\Bigl(
[S^{2}]_{\nu}^{\rho}[S^{2}]^{\sigma}_{\mu}+[S^{2}]_{\nu}^{\sigma}[S^{2}]_{\mu}^{\rho}
\Bigr) \\[1pt]
&
+c_{4}\,[S^{2}]_{\mu\nu}[S^{2}]^{\rho\sigma}
+c_{1}\,\Bigl([S^{3}]_{\mu\nu}S^{\rho\sigma}+S_{\mu\nu}[S^{3}]^{\rho\sigma}\Bigr)
+c_{2}\Bigl([S^{3}]_{\mu\nu}[S^{2}]^{\rho\sigma}+[S^{2}]_{\mu\nu}[S^{3}]^{\rho\sigma}\Bigr)
+ c_{3}\,[S^{3}]_{\mu\nu}[S^{3}]^{\rho\sigma} \,\Bigr] \,,
\end{aligned}
\ee
where the coefficients $c_i$ are given by
\begin{eqnarray}
c_{1} = \dfrac{e_{3}-e_{1}e_{2}}{-e_{1}e_{2}e_{3}+e_{3}^{2}+e_{1}^{2}e_{4}} \,,\;
 c_{2} = \dfrac{e_{1}^{2}}{-e_{1}e_{2}e_{3}+e_{3}^{2}+e_{1}^{2}e_{4}} \,, \;
c_{3} = \dfrac{-e_{1}}{-e_{1}e_{2}e_{3}+e_{3}^{2}+e_{1}^{2}e_{4}} \,,\;
c_{4} = \dfrac{e_{3}-e^3_{1}}{-e_{1}e_{2}e_{3}+e_{3}^{2}+e_{1}^{2}e_{4}} \,,\;
\end{eqnarray}
\end{widetext}
and here and henceforth all indices are moved with the metric $\gmn$.
Obviously, expression (\ref{d_Sm}) makes sense only if 
$-e_{1}e_{2}e_{3}+e_{3}^{2}+e_{1}^{2}e_{4}$ does not vanish, which in turn can be shown to be equivalent to the nonintersection of the spectra of $S^\mu_{\hphantom{\mu} \nu}$ and $-S^\mu_{\hphantom{\mu} \nu}$ mentioned above.

This result was checked to agree with the one we know from the mathematical literature \cite{Sylvesterpoly} using in particular non trivial identities -syzygies- which 
also play a fundamental role for the derivation of the covariant constraints below. These identities, as a consequence of the ``second fundamental theorem" of invariant theory \cite{Procesi}, can be derived using the Cayley-Hamilton theorem stating that for an arbitrary $4\times 4$ matrix $M$, one has 
\be \label{CHM}
M^{4} = e_{1}(M)  M^{3}-  e_{2}(M)M^{2} +  e_{3}(M) M - e_{4}(M)\mathbb{1}. 
\ee
One can then apply this to a matrix $M$ built out of 4 arbitrary matrices $A,B,C,D$ and four arbitrary real numbers $\{x_i\}$ in the form, $
M=x_0 A+x_1 B+x_2C+x_3D\,. $
Now, because the $\{x_i\}$ as well as the matrices $A,B,C,D$ are arbitrary, it means that in equation (\ref{CHM}) the terms which have the same degree of homogeneity in the $\{x_i\}$ must each yield separate identities between the matrices $A,B,C,D$. Once these identities are obtained, one can replace in them $A$ by $h$, $B$ by $S$, $C$ by $S^2$ and $D$ by $S^3$ to get non trivial matrix syzygies, denoted here as  $[{\cal I}_{k}]^\mu_{\hphantom{\mu}\nu}=0$, between the tensors of interest here. Notice that the above Cayley-Hamilton equation (\ref{CHM}) can also be used iteratively, when applied to the matrix $S$, to replace any power of $S$, $S^k$, with $k \geq 4$ by a linear combination of powers of $S$, $S^i$ with $i \leq 3$. This was done systematically in order to reach the expression (\ref{d_Sm}).

Using (\ref{d_Sm}), one can get the linearization of field equations (\ref{g_eom1_first}) around an arbitrary metric $g_{\mu \nu}$ reading~\footnote{Note that Guarato \& Durrer \cite{Guarato:2013gba} also obtained the quadratic action of dRGT theories around an arbitrary metric using different techniques from ours. Their analysis relied on being able to put $S$ in triangular form, which is only possible provided $S^2=g^{-1}f$ has no negative eigenvalues and if one consistently treats $\fmn$ as a tensor under general coordinate transformations. This however does not cover all possible real matrices $g^{-1}f$ with a well-defined real square root (see e.g.~\cite{Deffayet:2012zc}). Moreover, Ref.~\cite{Guarato:2013gba} then specialized to FLRW backgrounds using metrics which are not actual solutions of the background equations (see~\cite{Bernard:2015mkk}).}. 
\be \label{LinField}
\delta E_{\mu\nu} \equiv \delta\mathcal{G}_{\mu\nu}
+m^2\,\mathcal{M}_{\mu\nu}^{\ph\mu\ph\nu\rho\sigma}h_{\rho\sigma}=0\,.
\ee
where $\delta\mathcal{G}_{\mu\nu}$ is the linearization of the Einstein tensor
\ba
\delta\mathcal{G}_{\mu\nu}&=&-\tfrac1{2}\left[\delta^\rho_\mu\delta^\sigma_\nu\nabla^2+g^{\rho\sigma}\nabla_\mu\nabla_\nu 
-\delta^\rho_\mu\nabla^\sigma\nabla_\nu\right. \nonumber \\ && \left.
-\delta^\rho_\nu\nabla^\sigma\nabla_\mu 
 -g_{\mu\nu} g^{\rho\sigma}\nabla^2 +g_{\mu\nu}\nabla^\rho\nabla^\sigma\right]h_{\rho\sigma} \nonumber \\
&& 
+\tfrac1{2}\left[g_{\mu\nu} R^{\rho\sigma}-\delta^\rho_\mu \delta^\sigma_\nu R\right]h_{\rho\sigma}
\ea
and the ``mass matrix" $\mathcal{M}_{\mu\nu}^{\ph\mu\ph\nu\rho\sigma}$ is defined through
\be
\mathcal{M}_{\mu\nu}^{\ph\mu\ph\nu\rho\sigma}\equiv\frac{\p\Vmn}{\p g_{\rho\sigma}}\,,
\ee
and is given by the following expression,
\begin{widetext}
\ba\label{Mgen}
\nonumber
\mathcal{M}_{\mu\nu}^{\ph\mu\ph\nu\rho\sigma} &=&
 \frac1{2}V_{\mu}^{\ \sigma}\delta_{\nu}^{\rho} - \frac1{2}\left(\beta_2\delta_{\mu}^{\lambda}+\beta_3\left(e_1\delta_{\mu}^{\lambda}-S_{\mu}^{\lambda}\right)\right)[S^2]_{\lambda}^{\sigma}\,\delta_{\nu}^{\rho} 
+  \frac{1}{4}\,\sum_{n=1}^3\sum_{k=1}^n\sum_{m=1}^k(-1)^{n+k+m}\beta_n\,e_{k-m}
[S^{n-k}]^\lambda_{\ph\lambda\mu}\,g_{\nu\lambda}\,g^{\tau(\rho}\,[S^m]^{\sigma)}_{\ph\sigma\tau} \\
  &&-\frac1{2}\left(\beta_1\,\delta^{\tau}_{\ph\tau\lambda}+\beta_2\,e_1\,\delta^{\tau}_{\ph\tau\lambda}+\beta_3\left(e_2\,\delta^{\tau}_{\ph\tau\lambda}+ [S^2]^{\tau}_{\ph\tau\lambda}\right)\right)\dfrac{\delta S^{\lambda}_{\ph\lambda\mu}}{\delta g_{\rho\sigma}}\,g_{\nu\tau} + (\mu\leftrightarrow\nu) \,
\ea
\end{widetext}
Note that here, for the sake of simplicity, we have only considered the case of a massive graviton in vacuum, but it can easily be coupled to some matter source by introducing an energy-momentum tensor $\delta T_{\mu \nu}$ on the right-hand side of Eq.~(\ref{LinField}). Provided that this tensor is conserved with respect to the background metric $g_{\mu \nu}$ (as it is usually assumed in massive gravity, and follows from dRGT action if one just couples matter minimally to the dynamical metric), the results presented in the following on the presence of the extra scalar constraint will hold, since this follows from the linearized Bianchi identity.


In the most general case, Eq.~(\ref{LinField}) still contains two different metrics, the background dynamical metric $g_{\mu \nu}$ and the nondynamical metric $f_{\mu \nu}$  which can be traded for the tensor $S^2$. However, there is a subclass of dRGT models where one can explicitly get rid of this second metric and obtain field equations for a massive graviton on the background of just one metric $\gmn$. This subclass of models (called here $\beta_1$ models) is defined by setting to zero the parameters $\beta_2$ and $\beta_3$. In this case, indeed, $V_{\mu \nu}$ is linear in $S$ and the background field equations can just be reexpressed as \cite{Hassan:2013pca}
\be\label{b1_Ssol}
S^\rho_{~\nu}=\frac1{\beta_1m^2}\left[R^\rho_{~\nu}-\frac1{6}\delta^\rho_{\nu} R
-\frac{m^2\beta_0}{3}\,\delta^\rho_{\nu}\right]\,,
\ee
where $R_{\rho \nu}$ is the Ricci tensor of the metric $\gmn$, and $R$ the corresponding Ricci scalar. Using (\ref{Sdef}), this can equivalently be seen (by squaring the above equation) as expressing $\fmn$ in terms of $\gmn$ and its Ricci curvatures. This remarkable feature means that in the linearized equations of motion, we can eliminate any and all occurrences of the auxilliary metric $\fmn$ in favor of $\gmn$ and its curvature. This feature only requires a non-vanishing $\beta_1$~\footnote{We keep otherwise $\beta_0$ and $\beta_1$ arbitrary which means that these models, if one sets there $\fmn$ equal to the flat space-time metric $\eta_{\mu \nu}$, will not always have $\gmn=\eta_{\mu\nu}$ as a background solution, and hence will not always be nonlinear extensions of Fierz-Pauli theories {\it stricto sensu}. It is however always possible to get a flat space-time solution for $\gmn$ by adding a constant conformal factor in $\fmn$.}.

Having carried out this elimination, we take the obtained linearized field equation (\ref{LinField}) as a new starting point, and ask if one can show from these equations that the graviton $h_{\mu \nu}$ propagates 5 polarizations (or less) for a completely generic metric $\gmn$ (i.e.~without assuming it obeys the background equations). The idea here is to try to parallel what can be done for a massive graviton on flat space-time 
with metric $\eta_{\mu \nu}$ (and easily extended to a massive graviton on an Einstein space-time \cite{Higuchi:1986py,Buchbinder:1999ar}). 
In this case, the linearized Bianchi identities lead, by taking one derivative of the field equations, to four constraints reading 
\be \label{flatvect}
 \partial^\mu h_{\mu \nu} - \partial_\nu h = 0. 
\ee
Taking another derivative of this equation and subtracting this from the trace (tracing with $\eta_{\mu \nu}$) of the field equations, one then concludes that $h$ , defined as  $h = \eta^{\mu \nu} h_{\mu \nu}$, vanishes in vacuum.  Together with (\ref{flatvect}) this gives five Lagrangian constraints, which eliminate as many degrees of freedom out of the {\it a priori} 10 dynamical degrees of freedom of $h_{\mu \nu}$.

Let us then try to follow a similar path from the field equations (\ref{LinField}). First, it is easy to find four vector constraints similar to (\ref{flatvect}). Indeed, as a consequence of the Bianchi identities, one has 
\be
\nabla^\mu {\delta {\cal G}_{\mu \nu}} \sim 0
\ee
where here and henceforth $\nabla$ denotes the covariant derivative taken with respect to the background metric and 
two expressions separated by the symbol  ``$\sim$" are by definition equal {\it off shell} (i.e.~without using the field equations) up to terms containing no second or higher order derivatives acting on $h_{\mu \nu}$. 
Hence, the field equations yield the four vector constraints (being first order in derivatives)
\be \label{vectconsgen}
\nabla^\mu \delta E_{\mu\nu} =0.
\ee
In analogy with the flat space case we are interested in finding a fifth scalar constraint which generalizes the constraint $h=0$.
This fifth constraint should be the one which eliminates the Boulware-Deser ghost and reduces the number of degrees of freedom from 6 to 5. Accordingly, we look for a linear combination of scalars made by tracing over the field equations (\ref{LinField}), and its second derivatives, which would not contain any derivatives of $h_{\mu \nu}$ of order strictly higher than one. However, we have now at hand two (symmetric) tensors which can be used to take traces, namely the metric $\gmn$ and its Ricci curvature $R_{\mu \nu}$. Equivalently we can also use the metric and the tensor $S_{\mu \nu}$ trading $R_{\mu \nu}$ for $S_{\mu \nu}$ via equation (\ref{b1_Ssol}). Choosing the second solution turns out to be more convenient for technical reasons. We stress however that the two possibilities are strictly equivalent and do not impose any restriction on $\gmn$, since (\ref{b1_Ssol}) can just be considered as a definition of $S_{\mu \nu}$ in terms of $R_{\mu \nu}$ and $\gmn$ (as opposed to a background field equation).  
Hence we define the scalars $\Phi_i$ obtained by tracing the equations of motion with powers of $S$,
\be \label{defphi}
\Phi_i\equiv[S^i]^{\mu\nu}\,\delta E_{\mu\nu} \,,
\ee
together with the scalars $\Psi_i$ obtained by tracing the derivative of the divergence of the equations of motion in various ways,
\be \label{delpsi}
\Psi_{i}\equiv\frac{1}{2}\,[S^i]^{\mu \nu} \nabla_{\nu} \nabla^{\lambda}\,\delta E_{\lambda\mu}\,.
\ee 
An exhaustive set of linearly independents scalar is obtained by restricting $i$, $0\leq i\leq 3$, due the the Cayley-Hamilton identity.
To summarize we look for  a specific linear combination
 of the scalars $\Phi_{i}$ and $\Psi_{i}$, $i=0,..,3$ with scalar coefficients $\{\ccc_i,\ddd_i\}$ to be determined, such that 
\be \label{lincombwesearch}
\sum_{i=0}^{3}\left(\ccc_{i}\,\Phi_{i} + \ddd_{i}\,\Psi_{i}\right) \sim 0,
\ee
 i.e.~which contains no second (or higher) derivatives of $h_{\mu \nu}$.
Computing explicitly the scalars $\Phi_{i}$ and $\Psi_{i}$ \cite{Bernard:2015mkk}, one obtains that, in these scalars, the second derivatives of $h_{\mu \nu}$ appear in the form of 
linear combinations (with $S$-dependent coefficients) of 26 different scalars $\scah_i$ made by contracting $\nabla_\mu \nabla_\nu h_{\rho \sigma}$ with powers of $S$ (including zeroth power which is simply the metric) in various ways. Two of these scalars are e.g. $\Jaa$ and $\Abb$.

We get {\it a priori} 26 equations for the seven unknowns $\{\ccc_i,\ddd_i\}$~\footnote{There are eight scalars $\{\ccc_i,\ddd_i\}$, but they only need to be determined up to an overall factor.} by setting to zero each coefficient of the  $\scah_i$ which appears in (\ref{lincombwesearch}). However, one can show that not all the scalars  $\scah_i$ are independent, thanks to the syzygies ${\cal I}_k$. 
Indeed, the equation $\nabla_\mu \nabla^\nu[{\cal I}_{k}]^\mu_{\hphantom{\mu}\nu}=0$, together with the use of the Cayley-Hamilton theorem for $S$ yields four independent identities between the scalars $\scah_i$  which vanish up to terms $\sim 0$. These identities are just enough to reduce to seven the number of equations to be solved to fulfill (\ref{lincombwesearch}). This yields a unique solution for the coefficients $\{\ccc_i,\ddd_i\}$ which translates into the identity 
\begin{equation}
\dfrac{m^2\,\beta_1\,e_4}{4}\,\Phi_0 \sim - e_3\,\Psi_0 + e_2\,\Psi_1-e_1\,\Psi_2 + \Psi_3  \,.
\end{equation}
 Hence, using the field equations, we get the scalar constraint
\begin{equation} \label{scafin}
-\dfrac{m^2\,\beta_1\,e_4}{4}\,\Phi_0 - e_3\,\Psi_0 + e_2\,\Psi_1-e_1\,\Psi_2 + \Psi_3 = 0 \,,
\end{equation}
valid now for an arbitrary metric $\gmn$. Notice that the curvature of this metric enters this constraint in a nontrivial way via the tensor $S$.
One can check in fact that to the lowest nontrivial order (i.e.~at linear order) in curvature, the analysis done here agrees with the one of \cite{Buchbinder:1999ar} which investigates at this order the consistent coupling of a massive graviton to a curved background.

A nontrivial check is obtained considering the case where the background metric is covering an Einstein space-time with a cosmological constant $\Lambda$. In this case one has $R_{\mu \nu} = \Lambda g_{\mu \nu}$ and the constraint (\ref{scafin}) has the usual form found in Einstein space \cite{Higuchi:1986py,Buchbinder:1999ar}, i.e.~it reads 
\begin{equation}
h\left(1-\frac{2 \Lambda}{3 m^2_{FP}}\right) = 0, 
\end{equation}
where $m_{FP}$ is the graviton mass as appearing in the Fierz-Pauli action extended to Einstein space-time backgrounds. This mass is given here by the relation $3 m_{FP}^2 = \Lambda - m^2 \beta_0$. 
For generic Einstein space-times, including flat ones, this constraint reduces to $h=0$ which also shows that (\ref{scafin}) 
is independent of the vector constraints (\ref{vectconsgen}), as it should be. One also recovers the partially massless case (for $3 m^2_{FP} = 2 \Lambda$) where this constraint vanishes identically (i.e.~does not impose any constraint on the field $h_{\mu \nu}$) \cite{Deser:1983mm}.  In fact the current analysis can be used to further investigate the issue of partial masslessness on more general backgrounds. 

Another interesting direction to go would be to investigate whether an action for a single dynamical metric could be obtained leading to the linearized equations (\ref{LinField}). It appears however unlikely that such an action exists with a simple form given what we know from the dRGT theory.

\acknowledgments
The research of CD and MvS leading to these results have received
funding from the European Research Council under
the European Communitys Seventh Framework Programme
(FP7/2007-2013 Grant Agreement No. 307934).
In the process of checking our calculations, we have heavily used the XTENSOR package~\cite{xTensor}
developed by J.-M.~Mart\'{\i}n-Garc\'{\i}a for
MATHEMATICA. We thank Angnis Schmidt-May, Fawad Hassan and Kurt Hinterbichler for discussions.


\begin{thebibliography}{999}

\bibitem{reviews} 
  K.~Hinterbichler,
  Rev.\ Mod.\ Phys.\  {\bf 84}, 671 (2012)
  [arXiv:1105.3735 [hep-th]].
  C.~de Rham,
  Living Rev.\ Rel.\  {\bf 17}, 7 (2014)
  [arXiv:1401.4173 [hep-th]].

\bibitem{Fierz:1939ix}M. Fierz, Helv. Phys. Acta 12 (1939) 3;
M.~Fierz and W.~Pauli,
Proc.\ Roy.\ Soc.\ Lond.\ A {\bf 173}, 211 (1939).

\bibitem{Higuchi:1986py}
  A.~Higuchi,
  Nucl.\ Phys.\ B {\bf 282} (1987) 397.
  I.~Bengtsson,
  J.\ Math.\ Phys.\  {\bf 36}, 5805 (1995)
  [gr-qc/9411057].
  A.~Cucchieri, M.~Porrati and S.~Deser,
  Phys.\ Rev.\ D {\bf 51}, 4543 (1995)
  [hep-th/9408073].

  M.~Porrati,
  Phys.\ Lett.\ B {\bf 498}, 92 (2001)
  [hep-th/0011152].
  C.~Deffayet and S.~Randjbar-Daemi,
  Phys.\ Rev.\ D {\bf 84}, 044053 (2011)
  [arXiv:1103.2671 [hep-th]].


\bibitem{Buchbinder:1999ar}
  I.~L.~Buchbinder, D.~M.~Gitman, V.~A.~Krykhtin and V.~D.~Pershin,
  Nucl.\ Phys.\ B {\bf 584} (2000) 615
  [hep-th/9910188].
  I.~L.~Buchbinder, D.~M.~Gitman and V.~D.~Pershin,
  Phys.\ Lett.\ B {\bf 492}, 161 (2000)
  [hep-th/0006144].



\bibitem{deRham:2010kj} 
  C.~de Rham, G.~Gabadadze and A.~J.~Tolley,
  Phys.\ Rev.\ Lett.\  {\bf 106}, 231101 (2011)
  [arXiv:1011.1232 [hep-th]].
  C.~de Rham and G.~Gabadadze,
  Phys.\ Rev.\ D {\bf 82}, 044020 (2010)
  [arXiv:1007.0443 [hep-th]].
  C.~de Rham, G.~Gabadadze and A.~Tolley,
  arXiv:1107.3820 [hep-th].
  
\bibitem{Hassan:2011ea}
  S.~F.~Hassan and R.~A.~Rosen,
  arXiv:1111.2070 [hep-th].
  S.~F.~Hassan and R.~A.~Rosen,
  JHEP {\bf 1202} (2012) 126
  [arXiv:1109.3515 [hep-th]].
  S.~F.~Hassan, R.~A.~Rosen and A.~Schmidt-May,
  JHEP {\bf 1202} (2012) 026
  [arXiv:1109.3230 [hep-th]].
  S.~F.~Hassan and R.~A.~Rosen,
  Phys.\ Rev.\ Lett.\  {\bf 108} (2012) 041101
  [arXiv:1106.3344 [hep-th]].
  S.~F.~Hassan, A.~Schmidt-May and M.~von Strauss,
  Phys.\ Lett.\ B {\bf 715}, 335 (2012)
  [arXiv:1203.5283 [hep-th]].
	
	
\bibitem{Boulware:1973my} 
  D.~G.~Boulware and S.~Deser,
  Phys.\ Rev.\ D {\bf 6}, 3368 (1972).

\bibitem{defjac} 
  C.~Deffayet and T.~Jacobson,
  Class.\ Quant.\ Grav.\  {\bf 29}, 065009 (2012)
  [arXiv:1107.4978 [gr-qc]].


\bibitem{Hinterbichler:2012cn}
  K.~Hinterbichler and R.~A.~Rosen,
  JHEP {\bf 1207} (2012) 047
  [arXiv:1203.5783 [hep-th]].

\bibitem{Deffayet:2012nr}
  C.~Deffayet, J.~Mourad and G.~Zahariade,
  JCAP {\bf 1301} (2013) 032
  [arXiv:1207.6338 [hep-th]].

\bibitem{Deser:2014hga} 
  S.~Deser, M.~Sandora, A.~Waldron and G.~Zahariade,
  arXiv:1408.0561 [hep-th].

\bibitem{Banados:2013fda}
  M.~Banados, C.~Deffayet and M.~Pino,
  Phys.\ Rev.\ D {\bf 88} (2013) 12,  124016
  [arXiv:1310.3249 [hep-th]].

\bibitem{Sylvesterpoly}
Q.~Hu and D.~Cheng, App.\ Math.\  Lett.\, {\bf 19}, 9, 859 - 864 (2006). 

\bibitem{Bernard:2015mkk}
  L.~Bernard, C.~Deffayet and M.~von Strauss,
  arXiv:1504.04382 [hep-th].

\bibitem{Procesi}
C.~Procesi, Advances in Mathematics, {\bf 19}, 3, 306 - 381, 1976. 
G.~E.~Sneddon, J.\ Math.\ Phys.\ 39, 1659 (1998).

\bibitem{Guarato:2013gba}
  P.~Guarato and R.~Durrer,
  Phys.\ Rev.\ D {\bf 89} (2014) 084016
  [arXiv:1309.2245 [gr-qc]].
  
  \bibitem{Deffayet:2012zc}
  C.~Deffayet, J.~Mourad and G.~Zahariade,
  JHEP {\bf 1303} (2013) 086
  [arXiv:1208.4493 [gr-qc]].

\bibitem{Hassan:2013pca}
  S.~F.~Hassan, A.~Schmidt-May and M.~von Strauss,
  arXiv:1303.6940 [hep-th].

\bibitem{Deser:1983mm} 
  S.~Deser and R.~I.~Nepomechie,
  Annals Phys.\  {\bf 154}, 396 (1984).


\bibitem{xTensor}
J.-M.~Mart\'{\i}n-Garc\'{\i}a,
Comp. Phys. Commun. {\bf 179}, 597 (2008)
[arXiv:0803.0862 [cs.SC]],
$<$http://metric.iem.csic.es/Martin-Garcia/xAct/$>$.



 \end{thebibliography}
\end{document}